\documentclass[conference,a4paper]{IEEEtran}

\usepackage{etex}

\ifCLASSINFOpdf
\else
\fi

\usepackage{amsmath, amssymb, bm, cite, epsfig}
\usepackage{epstopdf}
\usepackage{graphicx}
\usepackage{array}
\usepackage{multirow}
\renewcommand{\arraystretch}{1}
\usepackage{amsfonts}
\usepackage{tabularx} 
\usepackage{tabu}
\usepackage{pbox}
\usepackage{makecell}

\usepackage{booktabs}
\usepackage{ctable}
\usepackage{color,soul}
\usepackage{xcolor,colortbl}
\definecolor{Gray}{gray}{0.85}
\definecolor{LightCyan}{rgb}{0.8,1,1}
\usepackage[utf8]{inputenc}
\usepackage[english]{babel}
 
\def\beq{\begin{equation}}
\def\eeq{\end{equation}}
\def\beqa{\begin{eqnarray}}
\def\eeqa{\end{eqnarray}}
\def\beqan{\begin{eqnarray*}}
\def\eeqan{\end{eqnarray*}}

\def\PL{\mathrm{PL}}
\def\dB{\mathrm{dB}}

\def\tm1{t\! - \! 1}
\def\tp1{t\! + \! 1}

\def\PL{\mathrm{PL}}
\def\dB{\mathrm{dB}}
\def\PLE{\mathrm{PLE}}
\def\FSPL{\mathrm{FSPL}}
\def\log{\mathrm{log}}
\def\FI{\mathrm{FI}}
\def\ABG{\mathrm{ABG}}
\def\CI{\mathrm{CI}}
\def\CIF{\mathrm{CIF}}

\usepackage{tikz}
\usetikzlibrary{calc}

\begin{document}
\title{Millimeter-Wave Distance-Dependent Large-Scale Propagation Measurements and Path Loss Models for Outdoor and Indoor 5G Systems}

%\author{\IEEEauthorblockN{Michael Shell}
%\IEEEauthorblockA{School of Electrical and\\Computer Engineering\\
%Georgia Institute of Technology\\
%Atlanta, Georgia 30332--0250\\
%Email: http://www.michaelshell.org/contact.html}
%\and
%\IEEEauthorblockN{Homer Simpson}
%\IEEEauthorblockA{Twentieth Century Fox\\

\author{\IEEEauthorblockN{Shu Sun$^*$, George R. MacCartney, Jr., and Theodore S. Rappaport}\\
\IEEEauthorblockA{NYU WIRELESS and Tandon School of Engineering, New York University, Brooklyn, NY, USA 11201\\
$^*$Corresponding author: ss7152@nyu.edu }

\thanks{Sponsorship for this work was provided by the NYU WIRELESS Industrial Affiliates program and NSF research grants 1320472, 1302336, and 1555332.}
}

\maketitle
\begin{tikzpicture}[remember picture, overlay]
\node at ($(current page.north) + (0,-0.25in)$) {S. Sun, G. R. MacCartney, Jr., and T. S. Rappaport, \rq\rq{}Millimeter-Wave Distance-Dependent Large-Scale Propagation Measurements};
\node at ($(current page.north) + (0,-0.4in)$) {and Path Loss Models for Outdoor and Indoor 5G Systems,\rq\rq{} in \textit{the 10th European Conference on Antennas and}};
\node at ($(current page.north) + (0,-0.55in)$) {\textit{Propagation (EuCAP 2016)}, April. 2016.};
\end{tikzpicture}

\begin{abstract}
This paper presents millimeter-wave propagation measurements for urban micro-cellular and indoor office scenarios at 28 GHz and 73 GHz, and investigates the corresponding path loss using five types of path loss models, the single-frequency floating-intercept (FI) model, single-frequency close-in (CI) free space reference distance model, multi-frequency alpha-beta-gamma (ABG) model, multi-frequency CI model, and multi-frequency CI model with a frequency-weighted path loss exponent (CIF), in both line-of-sight and non-line-of-sight environments. Results show that the CI and CIF models provide good estimation and exhibit stable behavior over frequencies and distances, with a solid physical basis and less computational complexity when compared with the FI and ABG models. Furthermore, path loss in outdoor scenarios shows little dependence on frequency beyond the first meter of free space propagation, whereas path loss tends to increase with frequency in addition to the increased free space path loss in indoor environments. Therefore, the CI model is suitable for outdoor environments over multiple frequencies, while the CIF model is more appropriate for indoor modeling. This work shows that both the CI and CIF models use fewer parameters and offer more convenient closed-form expressions suitable for analysis, without compromising model accuracy when compared to current 3GPP and WINNER path loss models. 
\end{abstract}

%\IEEEpeerreviewmaketitle
%\vspace{7pt}
\section{Introduction}\label{introduction}
The tremendous amount of raw available bandwidth at millimeter-wave (mmWave) frequency bands is an attractive resource to deliver multi-Gigabit-per-second (Gbps) data rates\cite{Rap13:Access,Rap15:TCOM}, and to relieve the mobile data traffic congestion in lower frequency bands, e.g., below 6 GHz. Fifth generation (5G) wireless communication systems that use mmWaves have been a research focus over the recent few years, and many research groups and industry entities have been conducting measurements and/or developing channel models at mmWave frequencies. For instance, outdoor propagation measurements and modeling at 60 GHz were carried out in various city streets\cite{Lov94, Smu97}; Aalto University performed outdoor channel measurements over 5 GHz of bandwidth from 81 GHz to 86 GHz in the E-band for point-to-point communications in a street canyon scenario in Helsinki, Finland\cite{Kyro12}; 28 GHz channel propagation measurements and models have been conducted by Samsung and NYU WIRELESS for 5G mobile communications\cite{Roh14,Rap15:TCOM,Mac15:Access}.

For both outdoor and indoor wireless communications networks, estimating large-scale path loss is important for modeling communications systems over distance and/or frequency, thus it is critical to accurately model path loss with users of the models in mind. When selecting models, those with fewer parameters that offer an intuitive or physics-based rationale, and that also have convenient closed-form expressions and provide repeatability and stability of parameters across many different data sets should be preferred over more complicated models that stray from the physics of propagation and offer widely varying model parameters when applied to different sets of data in similar physical environments.

Path loss models from empirical and simulated measurements are typically generated via two approaches:\cite{Rap15:TCOM}: ones that have some anchor based on the physics of transmitted power close to the antenna (such as the close-in (CI) free space reference distance model\cite{Rap15:TCOM}), and ones that do a mathematical curve fitting over the data set without any physical anchor to the transmitted power (such as the floating-intercept (FI) model\cite{3GPP:25996,WINNER}). The propagation models for mmWave frequencies in the ITU-R P-series\cite{ITU-R} consider free space loss and various additional effects such as tropospheric scatter and gaseous absorption, which are already inherently included in the CI and FI models. 

In this paper, extensive propagation measurements for the urban micro-cellular (UMi) street canyan (SC) scenario and indoor office scenario in New York City at mmWave frequencies of 28 GHz and 73 GHz are presented. Using the measured data, path loss results are studied and compared with five types of path loss models, the single-frequency FI model, the single-frequency CI model, the multi-frequency alpha-beta-gamma (ABG) model, the multi-frequency CI model, and the multi-frequency CI model with a frequency-weighted path loss exponent (CIF). Note that each of the five path loss models has a single slope, and dual-slope models can be constructed as long as continuity is maintained at the breakpoint\cite{Feu94}, but only marginal improvement in standard deviations (less than a dB) is typically obtained at the expense of additional model parameters and complexity.

\section{Millimeter-wave Propagation Measurements}
\subsection{Outdoor Measurements}
In the summers of 2012 and 2013, two outdoor propagation measurement campaigns were conducted at 28 GHz and 73 GHz, respectively, in downtown Manhattan, New York, where more than 10,000 directional power delay profiles (PDPs) were recorded using similar 400 Megachips-per-second (Mcps) spread spectrum sliding correlator channel sounders and directional steerable horn antennas at both the transmitter (TX) and receiver (RX) to investigate mmWave channel characteristics in a dense UMi environment \cite{Rap13:Access, Mac14}. The measurement system provided an RF first null-to-null bandwidth of 800 MHz and multipath time resolution of 2.5 ns. With the measurement system, the total time to acquire a PDP (including recording and averaging 20 instantaneous PDPs) was 40.94 ms$\times$20 = 818.8 ms, where 40.94 ms was the time it took to record a single PDP capture for a particular antenna pointing direction\cite{Rap15:TCOM}.

For the 28 GHz measurements conducted in Manhattan, three TX locations (heights of 7 m and 17 m) and 27 RX locations (heights of 1.5 m) were selected\cite{Rap15:TCOM}; a pair of 24.5 dBi-gain steerable directional horn antennas was used at the TX and RX with 10.9$^{\circ}$ and 8.6$^{\circ}$ half-power beamwidths (HPBWs) in the azimuth and elevation planes, respectively. For nine out of the ten measurement sweeps for each TX-RX location combination (except two line-of-sight (LOS) RX locations), the RX antenna was sequentially swept over the entire azimuth plane in increments of one HPBW at elevation angles of 0$^{\circ}$ and $\pm20^{\circ}$ about the horizon, so as to measure contiguous angular snapshots of the channel impulse response over the entire 360$^{\circ}$ azimuth plane at the RX, while the TX antenna remained at a fixed azimuth and elevation angle. For the final (tenth) measurement sweep, the TX antenna was swept over the entire azimuth plane at a fixed elevation (-10$^{\circ}$), with the RX antenna at a fixed azimuth and elevation plane~\cite{Rap13:Access}. 

For the 73 GHz measurements, five TX locations (heights of 7 m and 17 m) and 27 RX locations were used, with RX antenna heights of 2 m (mobile scenario) and 4.06 m (backhaul scenario), yielding a total of 36 TX-RX location combinations for the mobile (access) scenario and 38 combinations for the backhaul scenario. A pair of 27 dBi-gain rotatable directional horn antennas with a HPBW of 7$^{\circ}$ in both the azimuth and elevation planes was employed at the TX and RX. For each TX-RX location combination, TX and RX antenna azimuth sweeps were performed in steps of 8$^{\circ}$ or 10$^{\circ}$ at various elevation angles. Additional measurement procedures, hardware specifications, and channel modeling results can be found in~\cite{Rap13:Access,Mac14,Rap15:TCOM, Rap15}.

\subsection{Indoor Measurements}
During the summer of 2014, indoor propagation measurements at 28 GHz and 73 GHz were conducted on the 9\textsuperscript{th} floor of 2 MetroTech Center in downtown Brooklyn, New York, using the same 400 Mcps broadband sliding correlator channel sounders described for the outdoor measurements, with slight differences. The main differences between the channel sounder systems were the TX output powers (lower for indoor measurements) and the use of widebeam TX and RX antennas indoors (15 dBi, 28.8$^\circ$ azimuth HPBW at 28 GHz, and 20 dBi, 15$^\circ$ azimuth HPBW at 73 GHz)~\cite{Mac15:Access}. The indoor environment consisted of a cubicle-farm layout with long corridors, hallways, and closed offices and labs. Five TX locations and 33 RX locations were used at 28 GHz and 73 GHz, resulting in 48 TX-RX location combinations measured for each band, with three-dimensional (3D) T-R separation distances ranging from 3.9 m to 45.9 m, where more than 14,000 PDPs were measured. Of the 48 identical combinations for 28 GHz and 73 GHz, 10 were for LOS and 38 were for non-line-of-sight (NLOS) environments. In order to emulate an indoor hotspot scenario, the TX antennas were placed 2.5 m high near the 2.7 m ceiling, and the RX antennas were placed 1.5 m above the floor (to imitate a human carrying a mobile device). Additional information can be found in~\cite{Mac15:Access}.

\section{Large-scale Path Loss Models}
The five types of large-scale path loss models introduced in Section~\ref{introduction} are considered and compared using the outdoor and indoor data sets described above and detailed in~\cite{Mac15:Access,Rap15:TCOM}. The equation for the single-frequency FI model is given by~\eqref{FI1}:
\begin{equation}\label{FI1}
\begin{split}
\PL^{\FI}(d)[\dB]=&10\alpha \log_{10}\left(d\right)+\beta+\chi_{\sigma}^{\FI}\text{,   where } d\geq\textrm{ 1 m}
\end{split}
\end{equation}

\noindent where $\PL^{\FI}(d)$ denotes the path loss in dB as a function of the 3D T-R separation distance $d$, $\alpha$ is a coefficient characterizing the dependence of path loss on distance, $\beta$ is a floating intercept in dB\footnote{In some of our previous publications, $\alpha$ denoted the floating intercept and $\beta$ represented the distance coefficient. The two notations are swapped here to keep consistent with the notations in the ABG model given by Eq.~\eqref{ABG1}.}, and $\chi_{\sigma}^{\FI}$ is the shadow fading (SF) standard deviation describing large-scale signal fluctuations about the mean path loss over distance. The FI model is used in the WINNER II and 3GPP channel models\cite{3GPP:25996,WINNER}, but it requires two model parameters ($\alpha$ and $\beta$) and does not consider a physically-based anchor to the transmitted power. 

The equation for the CI model is given by~\eqref{CI1}:
\begin{equation}\label{CI1}
\begin{split}
\PL^{\CI}(f,d)[\dB]=\FSPL(f, 1~\textrm{m})[\dB]+10n\log_{10}\left(d\right)+\\\chi_{\sigma}^{\CI}\text{,   where } d\geq\textrm{ 1 m}
\end{split}
\end{equation}

\noindent where $n$ denotes the single model parameter, the \textit{path loss exponent} (PLE), with 10$n$ describing path loss in dB in terms of decades of distances beginning at 1 m (making it very easy to compute power over distance), $d$ is the 3D T-R separation distance, and $\FSPL(f,1~\textrm{m})=20\log_{10}\left(\frac{4\pi f}{c}\right)$ denotes the free space path loss in dB at a T-R separation distance of 1 m at the carrier frequency $f$, where $c$ is the speed of light. Note that the CI model has an intrinsic frequency dependence of path loss embedded within the 1 m FSPL value, and it has only one parameter, PLE, to be optimized. Furthermore, the CI model is applicable to both single- and multi-frequency cases. Free space path loss in the first meter of propagation ranges between 32 and 72 dB from 1 to 100 GHz, where a substantial amount of path loss in a practical mmWave communication system occurs. This first meter of loss is captured in the FSPL term, and is treated separately from the PLE which characterizes loss at distances greater than 1 m\cite{Rap15:TCOM}.

The ABG model aims to model large-scale path loss as a function of frequency as well as distance, and is expressed as follows:
\begin{equation}\label{ABG1}
\begin{split}
\PL^{\ABG}(f,d)[\dB]=10\alpha \log_{10}\left(\frac{d}{1~\textrm{m}}\right)+\beta&\\+10\gamma \log_{10}\left(\frac{f}{1~\textrm{GHz}}\right)+\chi_{\sigma}^{\ABG}\text{,}&\\ \text{  where } d\geq\textrm{ 1 m and }f\geq\text{ 1 GHz}
\end{split}
\end{equation}

\noindent where $\PL^{\ABG}(f,d)$ denotes the path loss in dB over frequency and distance, $\alpha$ and $\gamma$ are coefficients showing the dependence of path loss on distance and frequency, respectively, $\beta$ is an optimized offset (floating) value for path loss in dB, $f$ is the carrier frequency in GHz, and $\chi_{\sigma}^{\ABG}$ is the SF standard deviation describing large-scale signal fluctuations. The coefficients $\alpha$, $\beta$, and $\gamma$ are optimized from closed-form solutions that minimize the SF standard deviation\cite{Mac15:Access}. 

The CIF model is given by Eq.~\eqref{CIF1}\cite{Mac15:Access}:
\begin{equation}\label{CIF1}
\begin{split}
\PL^{\CIF}(f,d)[\dB]=&\FSPL(f, 1~\textrm{m})[\dB]+\\
&10n\bigg(1+b\Big(\frac{f-f_0}{f_0}\Big)\bigg)\log_{10}\left(d\right)+\chi_{\sigma}^{\CIF} \text{,}
\\ &\text{ where } d\geq\textrm{ 1 m}
\end{split}
\end{equation}

\noindent where $n$ denotes the distance dependence of path loss (similar to the PLE in the CI model), and $b$ is a model parameter that captures the amount of linear frequency dependence of path loss about the weighted average of all frequencies considered in the model. The parameter $f_0$ is a reference frequency that is an input parameter computed from the measurement set used in forming the model, and serves as the balancing point for the linear frequency dependence of the PLE, which is computed by~\eqref{CIF_f_0}:
\begin{equation}\label{CIF_f_0}
f_0 = \frac{\sum_{k=1}^{K}f_k N_k}{\sum_{k=1}^{K}N_k}
\end{equation}
where $K$ is the number of unique frequencies, $N_k$ is the number of path loss data points corresponding to the $k^{th}$ frequency $f_k$, and $\chi_{\sigma}^{\CIF}$ in~\eqref{CIF1} is the zero-mean Gaussian random variable (in dB) that describes large-scale shadowing. Note that the calculated $f_0$ is rounded to the nearest integer in GHz in this work. The CIF model reverts to the CI model for the single frequency case (when $f_0$ is equal to the single frequency $f$) or when $b=0$.

The CI and CIF models are based on fundamental principles of wireless propagation, dating back to Friis and Bullington, where the PLE is tied to the actual transmitted power using a close-in FSPL value, without the use of a floating intercept, and offers insight into path loss based on the environment, having a value of 2 in free space as shown by Friis' model and a value of 4 for the asymptotic two-ray ground bounce propagation model\cite{Rappaport:Wireless2nd}. The 1 m reference distance is a suggested standard that ties the transmitted power or path loss to a convenient close-in distance of 1 m\cite{Rap15:TCOM}. Standardizing to a reference distance of 1 m makes comparisons of measurements and models simpler, and provides a standard definition for the PLE, while enabling intuition and rapid computation of path loss. Emerging mmWave mobile systems will have very few users within a few meters of the base station antenna, and close-in users in the near field will have strong signals that will be power-controlled, compared to typical users much farther from the transmitter such that any path loss error in the near-field (between 1 m and the Fraunhofer distance) will be much smaller than the dynamic range of signals experienced by users in a commercial system\cite{Mac15:Access}. Additionally, the 1 m CI model offers more accurate prediction of path loss beyond measurement ranges when compared to the FI and ABG models (perhaps due to the fact that a great deal of path loss is accurately captured in the first meter of propagation close to the transmitter) as shown in\cite{Tho16:VTC,Akd14,Rap15:TCOM}. 

While the ABG model offers some physical basis in the $\alpha$ term, being based on a 1 m reference distance similar to the $n$ term in~\eqref{CI1}, it departs from physics when introducing both an offset $\beta$ (which is an optimization parameter that is not physically-based), and a frequency weighting term $\gamma$ which has no proven physical basis, although recent indoor measurements show that the path loss increases with frequency across the mmWave band\cite{Deng15} (both $\beta$ and $\gamma$ are used for curve fitting, as was done for the $\alpha$ term in the WINNER floating-intercept (\textit{alpha-beta}, or AB) model\cite{WINNER,Rap15:TCOM,GRM13:Globecom}). It is noteworthy that the ABG model is identical to the CI model if we equate $\alpha$ in the ABG model in~\eqref{ABG1} with the PLE $n$ in the CI model in~\eqref{CI1}, $\gamma$ in~\eqref{ABG1} with the free space PLE of 2, and $\beta$ in~\eqref{ABG1} with $20\log_{10}(4 \pi\times10^9/c)$.

\section{Outdoor Propagation Path Loss Results}
Using the five large-scale propagation path loss models presented above and the outdoor measurement data at both 28 GHz and 73 GHz, path loss parameters are analyzed and compared. The single-frequency FI and CI model parameters at 28 GHz and 73 GHz for the UMi SC scenario are contained in Table~\ref{tbl:FI_CI} (for the purpose of comparing path loss models and saving space, only omnidirectional path loss data measured with vertically-polarized TX and RX antennas are included; information on directional path loss and other polarization scenarios, as well as published raw data can be found in\cite{Rap15:TCOM,Mac15:PL}). It can be observed from Table~\ref{tbl:FI_CI} that the CI model provides intuitive path loss model parameter values due to its physical basis, while the parameters in the FI model sometimes contradict fundamental principles. For example, for the UMi SC LOS environment at 73 GHz, the CI model generates a PLE of 2.0, which matches well with the theoretical free space PLE of 2; however, the $\alpha$ in the FI model is -0.8, meaning that the path loss \textit{decreases} with distance, which is obviously not reasonable or physically possible in a passive channel. 

The path loss results for the UMi SC scenario using both the 28 GHz and 73 GHz outdoor measurements data sets for the multi-frequency ABG, CI, and CIF models are provided in Table~\ref{tbl:ABG_CI_CIF}. As shown by Table~\ref{tbl:ABG_CI_CIF}, for the LOS environment, both the CI and CIF models provide a PLE or $n$ of 2.0, which agrees very well with the theoretical free space PLE of 2. In contrast, the ABG model yields an $\alpha$ of 1.0, substantially lower than the theoretical free space PLE, indicating its lack of physical intuition. Meanwhile, the SF standard deviations for the ABG, CI, and CIF models are virtually identical, with a maximum difference of only 0.2 dB. The CIF model yields a value of $n$ that is identical to the PLE in the CI model for the UMi SC scenario. The frequency term $b$ in the CIF model is very small, i.e., -0.06 and -0.00 in LOS and NLOS environments, respectively, indicating that path loss has negligible frequency dependence beyond the first meter of propagation in UMi channels at mmWave frequencies, thus proving that the single-parameter CI model may be used for LOS and NLOS outdoor channels.

\begin{table}
%\captionsetup{width=\textwidth}
\renewcommand{\arraystretch}{1.0}
\begin{center}
\caption{Parameters for the single-frequency FI and CI path loss models in UMi and indoor office scenarios. SC denotes street canyon, and Dist. Range denotes distance range.}~\label{tbl:FI_CI}
\scalebox{0.87}{
\begin{tabular}{|c|c|c|c|c|c|c|c|}
\hline 
 Sce. & Env.& \makecell{Freq.\\(GHz)}&\makecell{Dist. \\Range\\(m)}&Model & \makecell{$\PLE$\\/$\alpha$} & \makecell{$\beta$ \\($\dB$)} & \makecell{$\sigma$ \\($\dB$)} \\ \specialrule{1.5pt}{0pt}{0pt}
\multirow{8}{*}{\makecell{UMi\\SC}} & \multirow{4}{*}{LOS}& \multirow{2}{*}{28}& \multirow{2}{*}{31-54} & FI & 3.9 & 31.8 & 2.9 \\ \cline{5-8}
 & & & & CI & 2.1 & -  & 3.5 \\ \cline{3-8}
 & & \multirow{2}{*}{73}& \multirow{2}{*}{27-54} &  FI & -0.8 & 115.6  & 3.9 \\ \cline{5-8}
 & & & & CI & 2.0 & - & 4.9 \\ \cline{2-8}
 & \multirow{4}{*}{NLOS}  & \multirow{2}{*}{28}& \multirow{2}{*}{61-186} &  FI & 2.5 & 80.6 & 9.7 \\ \cline{5-8}
 & & & & CI & 3.4 & - & 9.7 \\ \cline{3-8}
 & & \multirow{2}{*}{73}& \multirow{2}{*}{48-190} &  FI & 2.9 & 80.6 & 7.8 \\ \cline{5-8}
 & & & & CI & 3.4 & - & 7.9 \\ \specialrule{1.5pt}{0pt}{0pt}
\multirow{8}{*}{\makecell{Indoor\\Office}} & \multirow{4}{*}{LOS}& \multirow{2}{*}{28}& \multirow{2}{*}{4.1-21.3} &  FI & 1.2 & 60.4 & 1.8 \\ \cline{5-8}
 & & & & CI & 1.1 & -  & 1.8 \\ \cline{3-8}
 & & \multirow{2}{*}{73}& \multirow{2}{*}{4.1-21.3} &  FI & 0.5 & 77.9 & 1.4 \\ \cline{5-8}
 & & & & CI & 1.3 & - & 2.4 \\ \cline{2-8}
 & \multirow{4}{*}{NLOS}  & \multirow{2}{*}{28}& \multirow{2}{*}{3.9-45.9} &  FI & 3.5 & 51.3 & 9.3 \\ \cline{5-8}
 & & & & CI & 2.7 & - & 9.6 \\ \cline{3-8}
 & & \multirow{2}{*}{73}& \multirow{2}{*}{3.9-41.9} &  FI & 2.7 & 76.3 & 11.2 \\ \cline{5-8}
 & & & & CI & 3.2 & - & 11.3 \\ \cline{1-8}
\end{tabular}}
\end{center}
\end{table} 

\begin{table}
\renewcommand{\arraystretch}{1.0}
\begin{center}
\caption{Parameters for the multi-frequency ABG, CI, and CIF path loss models in UMi SC and indoor office scenarios. SC denotes street canyon, Env. stands for environment, Dist. Range denotes distance range, L means LOS, and N denotes NLOS.}~\label{tbl:ABG_CI_CIF}
\scalebox{0.87}{
\begin{tabular}{|c|c|c|c|c|c|c|c|c|}
\hline 
 Sce. & Env.& \makecell{Freq.\\(GHz)}&\makecell{Dist. \\Range\\(m)}&Model & \makecell{$\PLE$\\/$\alpha$/$n$} & \makecell{$\beta$ \\($\dB$)} & \makecell{$\gamma$\\/$b$} & \makecell{$\sigma$ \\($\dB$)} \\ \specialrule{1.5pt}{0pt}{0pt}
 \multirow{6}{*}{\makecell{UMi\\SC}} & \multirow{3}{*}{L}& \multirow{3}{*}{\makecell{28,\\73.5}}& \multirow{3}{*}{27-54} & ABG & 1.0 & 55.0 & 1.7 & 4.3 \\ \cline{5-9}
 & & & & CI & 2.0 & - & - & 4.5 \\ \cline{5-9}
 & & & & CIF & 2.0 & - & -0.06 & 4.4 \\ \cline{2-9}
 & \multirow{3}{*}{N}  & \multirow{3}{*}{\makecell{28,\\73.5}}  & \multirow{3}{*}{48-190} & ABG & 2.8 & 46.7 & 1.9 & 8.4 \\ \cline{5-9}
 & & & & CI & 3.4 & - & - & 8.4 \\ \cline{5-9}
 & & & & CIF & 3.4 & - & -0.00 & 8.4 \\ \specialrule{1.5pt}{0pt}{0pt}
\multirow{6}{*}{\makecell{Indoor\\Office}} & \multirow{3}{*}{L}& \multirow{3}{*}{\makecell{28,\\73.5}}& \multirow{3}{*}{4.1-21.3} & ABG & 0.9 & 26.8 & 2.6 & 1.8 \\ \cline{5-9}
 & & & & CI & 1.2 & - & - & 2.3 \\ \cline{5-9}
 & & & & CIF & 1.2 & - & 0.18 & 2.1 \\ \cline{2-9}
 & \multirow{3}{*}{N}  & \multirow{3}{*}{\makecell{28,\\73.5}}  & \multirow{3}{*}{3.9-45.9} & ABG & 3.1 & 1.3 & 3.8 & 10.3 \\ \cline{5-9}
 & & & & CI & 2.9 & - & - & 10.9 \\ \cline{5-9}
 & & & & CIF & 3.0 & - & 0.21 & 10.4 \\ \cline{1-9}
\end{tabular}}
\end{center}
\end{table} 

\section{Indoor Propagation Path Loss Results}
In order to characterize co-polarization signal attenuation as a function of distance and frequency for the indoor office channel, the parameters for the single-frequency FI and CI models, and the multi-frequency ABG, CI, and CIF models are provided and compared, as previously published with raw data in~\cite{Mac15:Access}. The resulting single-frequency path loss model parameters emphasize the frequency dependence of indoor path loss beyond the first meter of FSPL, where PLEs at 73 GHz are larger than 28 GHz PLEs, as shown in Table~\ref{tbl:FI_CI}. Specifically, LOS PLEs are 1.1 and 1.3 at 28 GHz and 73 GHz, respectively, indicating constructive interference and waveguiding effects in LOS indoor channels at mmWave frequencies. Furthermore, the NLOS PLEs are 2.7 and 3.2 at 28 GHz and 73 GHz, respectively, showing that 73 GHz propagating waves attenuate by 5 dB more per decade of distance in the indoor environment beyond the first meter, as provided in Table~\ref{tbl:FI_CI}. The FI model indicates lower attenuation as a function of log-distance in some cases (73 GHz NLOS $\alpha$ = 2.7 compared to $n$ of 3.2, and 73 GHz LOS $\alpha$ = 0.5 compared to $n$ = 1.3), however, the FI model parameters can exhibit strange, non-physics based values, specifically $\alpha$ = 0.5 for 73 GHz LOS which implies ultra-low loss with distance (less than in a waveguide) that does not follow basic physics. The physically-based 1 m FSPL anchor of the CI model for single frequencies allows for a simpler model (only one parameter) with virtually no decrease in model accuracy (standard deviation of 9.3 dB and 9.6 dB for FI and CI, respectively in NLOS at 28 GHz, and 11.2 dB and 11.3 dB for FI and CI, respectively in NLOS at 73 GHz) by representing free space propagation close to the transmitting antenna.

\begin{figure}
\centering
 \includegraphics[width=2.7in]{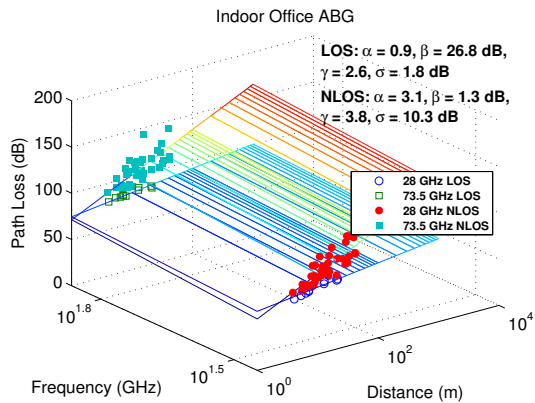}
    \caption{Multi-frequency ABG indoor path loss models across different frequencies and distances in LOS and NLOS environments.}
    \label{fig:Indoor_ABG}
\end{figure}

\begin{figure}
\centering
 \includegraphics[width=2.7in]{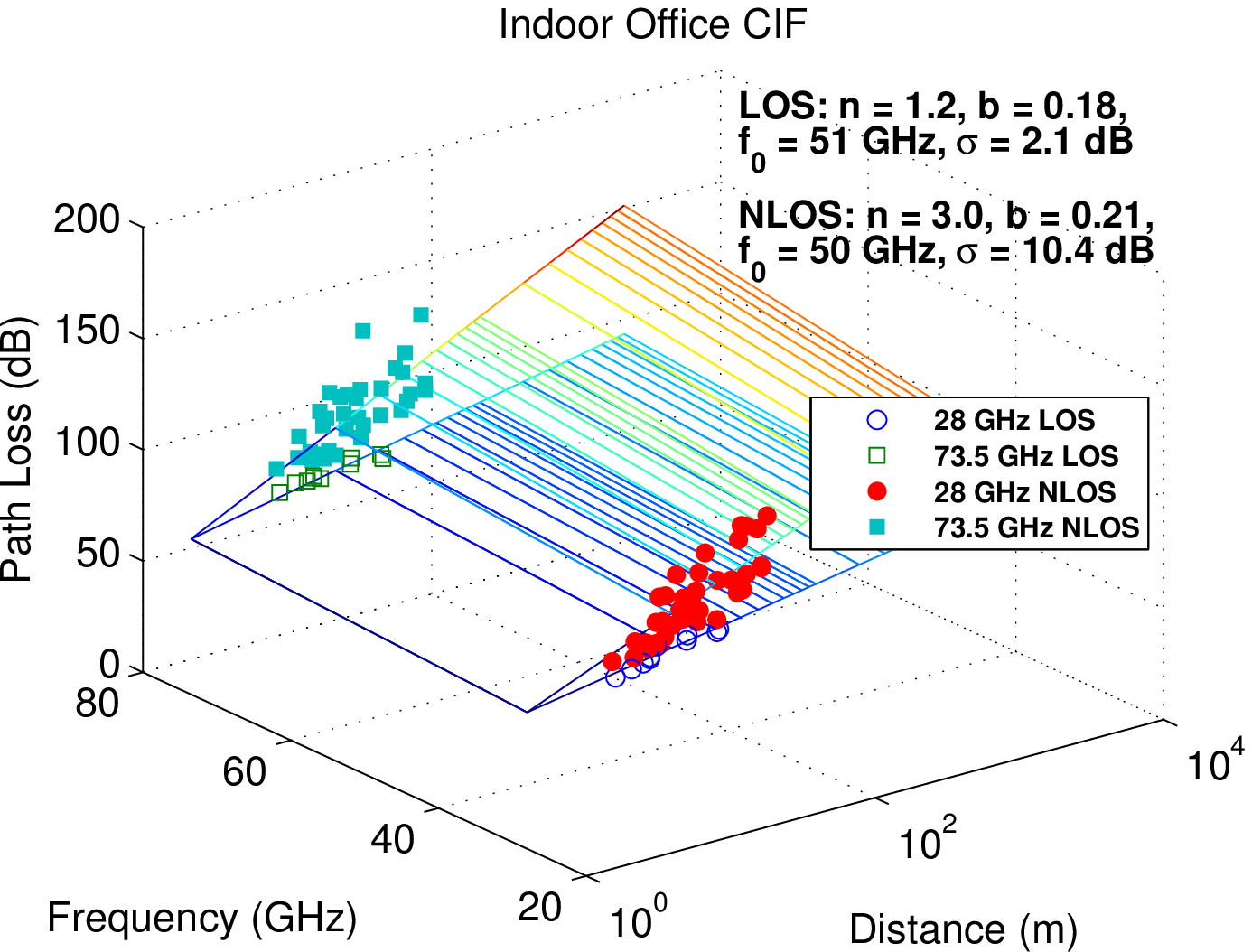}
    \caption{Multi-frequency CIF indoor path loss models across different frequencies and distances in LOS and NLOS environments.}
    \label{fig:Indoor_CIF}
\end{figure}

\begin{figure}
\centering
 \includegraphics[width=2.7in]{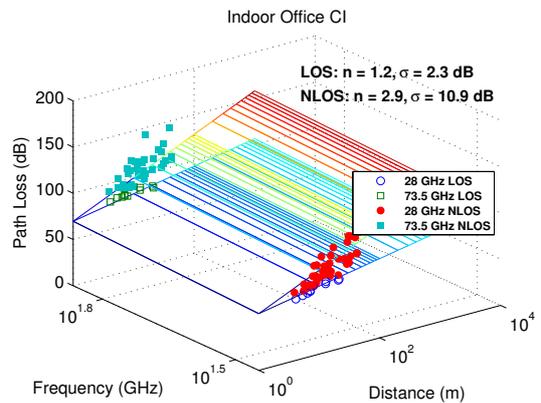}
    \caption{Multi-frequency CI indoor path loss models across different frequencies and distances in LOS and NLOS environments.}
    \label{fig:Indoor_CI}
\end{figure}

The multi-frequency path loss models (ABG, CIF, and CI) allow for the comparison of distance and frequency dependence at the 28 GHz and 73 GHz mmWave bands in the indoor office environment. Figs.~\ref{fig:Indoor_ABG},~\ref{fig:Indoor_CIF}, and~\ref{fig:Indoor_CI} show the omnidirectional path loss data in LOS and NLOS environments at 28 GHz and 73 GHz and the corresponding multi-frequency model parameters and fits to the data, while the model parameters are also provided in Table~\ref{tbl:ABG_CI_CIF}. Similar to the single-frequency CI model, the multi-frequency CI (still just one parameter) and CIF (only two parameters) models illustrate the physical basis via a free space reference distance at 1 m, while the CIF model includes a frequency-dependent balancing term $b$. The added benefit of the frequency-dependent term in the CIF model is the improvement in model accuracy, i.e. reduction in standard deviation (2.3 dB (CI) compared to 2.1 dB (CIF) in LOS, and 10.9 dB (CI) compared to 10.4 dB (CIF) in NLOS), as the CIF model has a better fit to the indoor data than the CI model, inherent in the frequency dependence of path loss observed in indoor environments. The CI PLE and CIF $n$ parameters are identical (1.2) in LOS and are extremely close (CI PLE of 2.9 and CIF $n$ of 3.0) in NLOS, adding credence to their physical significance and stability. The ABG model (with three parameters) provides slightly lower standard deviations in both LOS and NLOS environments compared to the CI and CIF models, but as was seen in the single-frequency FI model, ABG model parameters result in a slope attenuation term of 0.9 in LOS environments over log-distance (less loss than a coaxial transmission line or waveguide), thus indicating an ultra-low loss channel that lacks agreement with the fundamental physics of propagation. Furthermore, the very slight differences in path loss model standard deviations in NLOS environments (10.3 dB for ABG, 10.4 dB for CIF, and 10.9 dB for CI) are virtually indiscernible, well within measurement error and well within an order of magnitude for standard deviations that are above 10 dB, and reasonably within measurement error. Therefore, the more physically-sound and simpler CI and CIF models with a 1 m free space reference distance term are more convenient to model indoor mmWave channels.

\section{Comparison of Outdoor and Indoor Path Loss Results}
By comparing the multi-frequency path loss model parameters between the UMi SC scenario and indoor office scenario in Table~\ref{tbl:ABG_CI_CIF}, it can be found that the magnitude of $b$ in the CIF model is generally smaller for outdoor environments (-0.06 and -0.00 for LOS and NLOS, respectively) than for indoor environments (0.18 and 0.21 for LOS and NLOS, respectively), indicating less (in fact no frequency dependence whatsoever when $b$=0) frequency dependence of path loss in outdoor environments, beyond the first meter of free space propagation. The CIF model is quite similar to the CI model in terms of model parameters for outdoor environments and the extra frequency-dependent term of the CIF model is not needed, since the first meter of propagation captures virtually all of the frequency-dependent loss. We conclude that the CI model is thus most suitable for outdoor mmWave environments, as compared to the ABG and CIF models, while the CIF model is preferable to ABG and CI for indoor environments. 

Furthermore, comparing the outdoor and indoor LOS PLEs for the multi-frequency CI model in Table~\ref{tbl:ABG_CI_CIF}, it is observed that the outdoor LOS PLE (2.0) agrees well with the theoretical free space PLE of 2, whereas the indoor LOS PLE (1.2) is much lower than 2, due to waveguiding effects that enhance the received signal strength in indoor office environments. In addition, the NLOS multi-frequency PLE is also smaller for indoor environments (2.9) than outdoor (3.4), which may be ascribed to waveguiding effects and more strong reflected paths in the indoor office environments.

\section{Conclusion}
This paper describes the mmWave propagation measurements in both UMi SC and indoor office scenarios at both 28 GHz and 73 GHz, and presents and compares the single-frequency FI and CI path loss models, as well as the multi-frequency ABG, CI, and CIF models, using the data from extensive measurement campaigns. Single-frequency path loss results show that the CI model is preferable compared to the FI model (presently used in WINNER and 3GPP) for both outdoor and indoor environments, due to its physical basis, simplicity, and robustness over measured frequencies and distance ranges. Multi-frequency analysis shows that the CI model is suitable for outdoor environments because of its physical basis, stability, simplicity, and the fact that measured path loss exhibits little dependence on frequency in outdoor environments, beyond the first meter of free space propagation (that is captured in the CI model). On the other hand, the CIF model is well suited for indoor environments, since it is based on physics, requires only two parameters as a natural extension of the CI model, and incorporates the frequency dependence feature of path loss observed in indoor environments.

%\section*{Acknowledgment}
%Sponsorship for this work was provided by the NYU WIRELESS Industrial Affiliates program and NSF research grants 1320472, 1302336, and 1555332.

\ifCLASSOPTIONcaptionsoff
  \newpage
\fi

\bibliographystyle{IEEEtran}
\bibliography{bibliography}

\begin{thebibliography}{10}
\providecommand{\url}[1]{#1}
\csname url@samestyle\endcsname
\providecommand{\newblock}{\relax}
\providecommand{\bibinfo}[2]{#2}
\providecommand{\BIBentrySTDinterwordspacing}{\spaceskip=0pt\relax}
\providecommand{\BIBentryALTinterwordstretchfactor}{4}
\providecommand{\BIBentryALTinterwordspacing}{\spaceskip=\fontdimen2\font plus
\BIBentryALTinterwordstretchfactor\fontdimen3\font minus
  \fontdimen4\font\relax}
\providecommand{\BIBforeignlanguage}[2]{{%
\expandafter\ifx\csname l@#1\endcsname\relax
\typeout{** WARNING: IEEEtran.bst: No hyphenation pattern has been}%
\typeout{** loaded for the language `#1'. Using the pattern for}%
\typeout{** the default language instead.}%
\else
\language=\csname l@#1\endcsname
\fi
#2}}
\providecommand{\BIBdecl}{\relax}
\BIBdecl

\bibitem{Rap13:Access}
\text{T. S. Rappaport} \textit{et al.}, ``Millimeter wave mobile communications
  for \text{5G} cellular: It will work!'' \emph{IEEE Access}, vol.~1, pp.
  335--349, 2013.

\bibitem{Rap15:TCOM}
T.~S. Rappaport \emph{et~al.}, ``Wideband millimeter-wave propagation
  measurements and channel models for future wireless communication system
  design ({I}nvited {P}aper),'' \emph{IEEE Transactions on Communications},
  vol.~63, no.~9, pp. 3029--3056, Sep. 2015.

\bibitem{Lov94}
\text{G. Lovnes} \textit{et al.}, ``Channel sounding measurements at 59 ghz in
  city streets,'' in \emph{5th IEEE International Symposium on Personal, Indoor
  and Mobile Radio Communications}, vol.~2, Sep. 1994, pp. 496--500.

\bibitem{Smu97}
P.~Smulders and L.~Correia, ``Characterisation of propagation in 60 {GHz} radio
  channels,'' \emph{Electronics Communication Engineering Journal}, vol.~9,
  no.~2, pp. 73--80, Apr. 1997.

\bibitem{Kyro12}
M.~Kyro \emph{et~al.}, ``Experimental propagation channel characterization of
  mm-wave radio links in urban scenarios,'' \emph{IEEE Antennas and Wireless
  Propagation Letters}, vol.~11, pp. 865--868, 2012.

\bibitem{Roh14}
\text{W. Roh} \textit{et al.}, ``Millimeter-wave beamforming as an enabling
  technology for 5{G} cellular communications: theoretical feasibility and
  prototype results,'' \emph{IEEE Communications Magazine}, vol.~52, no.~2, pp.
  106--113, Feb. 2014.

\bibitem{Mac15:Access}
G.~R. {MacCartney, Jr.}, T.~S. Rappaport, S.~Sun, and S.~Deng, ``Indoor office
  wideband millimeter-wave propagation measurements and channel models at 28
  {GHz} and 73 {GHz} for ultra-dense 5{G} wireless networks ({I}nvited
  {P}aper),'' \emph{IEEE Access}, pp. 2388--2424, Dec. 2015.

\bibitem{3GPP:25996}
\text{3GPP TR 25.996}, ``Spatial channel model for multiple input multiple
  output \text{(MIMO)} simulations,'' Sep. 2012.

\bibitem{WINNER}
\text{P. Kyosti, \textit{et al.}}, ``\text{WINNER II channel models},''
  \emph{European Commission, \text{IST-4-027756-WINNER, Tech. Rep. D1.1.2}},
  Sep. 2007.

\bibitem{ITU-R}
\text{ITU-R P.620-6}, ``Propagation data required for the evaluation of
  coordination distances in the frequency range 100 {MHz} to 105 {GHz},'' 2005.

\bibitem{Feu94}
M.~Feuerstein \emph{et~al.}, ``Path loss, delay spread, and outage models as
  functions of antenna height for microcellular system design,'' \emph{IEEE
  Transactions on Vehicular Technology}, vol.~43, no.~3, pp. 487--498, Aug.
  1994.

\bibitem{Mac14}
G.~MacCartney and T.~Rappaport, ``73 \text{GHz} millimeter wave propagation
  measurements for outdoor urban mobile and backhaul communications in new york
  city,'' in \emph{2014 IEEE International Conference on Communications (ICC)},
  Jun. 2014, pp. 4862--4867.

\bibitem{Rap15}
T.~S. Rappaport, R.~W. {Heath, Jr.}, R.~C. Daniels, and J.~N. Murdock,
  \emph{Millimeter Wave Wireless Communications}.\hskip 1em plus 0.5em minus
  0.4em\relax Pearson/Prentice Hall 2015.

\bibitem{Rappaport:Wireless2nd}
T.~S. Rappaport, \emph{Wireless Communications: Principles and Practice},
  2nd~ed.\hskip 1em plus 0.5em minus 0.4em\relax Upper Saddle River, NJ:
  Prentice Hall, 2002.

\bibitem{Tho16:VTC}
\text{Timothy A. Thomas,~\textit{et al.}}, ``A prediction study of path loss
  models from 2-73.5 {GHz} in an urban-macro environment,''
  \emph{\textit{accepted to 2016 IEEE 83rd Vehicular Technology Conference (VTC
  Spring)}}, May 2016.

\bibitem{Akd14}
M.~R. Akdeniz \emph{et~al.}, ``Millimeter wave channel modeling and cellular
  capacity evaluation,'' \emph{IEEE Journal on Selected Areas in
  Communications}, vol.~32, no.~6, pp. 1164--1179, June 2014.

\bibitem{Deng15}
S.~Deng, M.~K. Samimi, and T.~S. Rappaport, ``28 \text{GHz} and 73 \text{GHz}
  millimeter-wave indoor propagation measurements and path loss models,'' in
  \emph{2015 IEEE International Conference on Communications (ICC), ICC
  Workshops}, Jun. 2015.

\bibitem{GRM13:Globecom}
\text{G. R. MacCartney, Jr.} \textit{et al.}, ``Path loss models for 5{G}
  millimeter wave propagation channels in urban microcells,'' in \emph{IEEE
  Global Communications Conference (GLOBECOM)}, Dec. 2013, pp. 3948--3953.

\bibitem{Mac15:PL}
G.~R. {MacCartney, Jr.} \emph{et~al.}, ``Millimeter-wave omnidirectional path
  loss data for small cell 5{G} channel modeling,'' \emph{IEEE Access}, vol.~3,
  pp. 1573--1580, Sept. 2015.

\end{thebibliography}

\end{document}